\documentclass[11pt]{article}

\usepackage{amsmath} 
\usepackage{cite}

\newcommand{\mycaption}[2][kurz]{{\begin{center} 
\parbox{15cm}{{\bf \caption[#1]{\rm {#2}}}} \end{center} }}
\newcommand{\pb}[3]{{\parbox{#1 cm}{ \vspace{#2 cm} \begin{center} #3 \end{center}}} 
\vspace{#2 cm}}

\newcommand{\la}{{\mathcal L}}

\newcommand{\equ}[1]{{(\ref{#1})}}

\newcommand{\sect}[1]{{Section \ref{#1}}}

\newcommand{\lr}[1]{{ \left( \, #1 \, \right) }}

\newcommand{\mr}[1]{{\mathrm{#1}}}

\begin{document}

\begin{flushright}
MC-TH-2002-09\\
WUE-ITP-2002-029\\
hep-ph/0210410\\
October
\end{flushright}

\begin{center}
{\large {\bf  LEP Constraints on 5-Dimensional Extensions of the
Standard  Model}}\\[0.5cm]
Alexander M\"uck$^{\,  a}$, Apostolos Pilaftsis$^{\, b}$\footnote{Talk
given at the 10th international conference on 
``Supersymmetry and Unification of Fundamental Interactions,''
Hamburg, Germany, 17--23 June 2002.} and Reinhold R\"uckl$^{\,a}$\\[0.2cm]    
$^a${\em Institut f\"ur Theoretische Physik und Astrophysik, 
Universit\"at  W\"urzburg,\\ 
Am Hubland,  97074 W\"urzburg,  Germany}\\[0.1cm] 
$^b${\em  Department of
Physics and Astronomy, University of Manchester,\\ 
Manchester M13 9PL,
United Kingdom}
\end{center}

\begin{center} 
\parbox{14cm}{
\vspace{0.2cm} \small  \centerline{\bf   ABSTRACT}  
We study  minimal 5-dimensional extensions of  the Standard  Model, in
which all or only some of the SU(2)$_L$  and U(1)$_Y$ gauge fields and
Higgs bosons  propagate in  the fifth compact  dimension.  In  all the
5-dimensional settings, the fermions are assumed to  be localized on a
3-brane.  In  addition,    we present  the  consistent  procedure  for
quantizing   5-dimensional   models    in the    generalized   $R_\xi$
gauge.  Bounds on the compactification   scale  between 4 and   6~TeV,
depending on   the  model, are   established by  analyzing electroweak
precision measurements and LEP2 cross sections. }
\end{center}

\setcounter{equation}{0}
\section{Introduction}

\indent

In  the  original  formulations  of  string theory~\cite{review},  the
compactification radius  $R$ of  the  extra dimensions and the  string
mass $M_s$ were considered to be  set by the 4-dimensional Planck mass
$M_{\rm P} = 1.9\times 10^{16}$~TeV.    However, recent studies   have
shown~\cite{IA,JL,EW,ADS,DDG} that  conceivable  scenarios of  stringy
nature  may exist for  which $R$ and   $M_s$ practically decouple from
$M_{\rm P}$.  For example, in the model  of Ref.~\cite{ADS}, $M_s$ may
become as low as a few TeV. In  this  case, $M_s$ constitutes  the
only fundamental scale in nature at which all forces including gravity
unify.  This low string-scale effective model could be embedded within
e.g.\ type I string theories~\cite{EW}, where  the Standard Model (SM)
may be  described  as   an  intersection of   higher-dimensional  $Dp$
branes~\cite{ADS,DDG,AB}.

As such  intersections may be higher  dimensional as well, in addition
to    gravitons   the  SM   gauge  fields   could    also    propagate  
within a higher-dimensional subspace with  compact dimensions of order
TeV$^{-1}$ for phenomenological  reasons. Since such low  string-scale
constructions may result in different higher-dimensional extensions of
the   SM~\cite{AB},     the actual    experimental   limits    on  the
compactification  radius  are,   to  some   extent,  model  dependent.
Nevertheless, most of   the  derived phenomenological limits  in   the
literature   were  obtained by assuming that  all the  SM gauge fields
propagate        in        a          common        higher-dimensional 
space~\cite{NY,WJM,CCDG,DPQ2,RW,DPQ1,CL}.

Here,   we wish  to  lift  the  above restriction   and  focus on  the
phenomenological consequences  of  models which  minimally depart from
the  assumption of a universal higher-dimensional scenario~\cite{MPR}.
Specifically, we  will  consider 5-dimensional  extensions   of the SM
compactified  on an  $S^1/Z_2$  orbifold,   where the  SU(2)$_L$   and
U(1)$_Y$ gauge bosons may not both live in the same higher-dimensional
space, the  so-called bulk.  In  all  our models, the SM  fermions are
localized on the  4-dimensional subspace, i.e.~on  a 3-brane or, as it
is   often called,  brane.    For each  higher-dimensional   model, we
calculate the   effects  of the  fifth  dimension  on  the electroweak
observables  and     analyze  their    impact  on    constraining  the
compactification scale.

The organization of this brief report is as follows:   in Section 2 we 
introduce  the  basic  concepts  of  higher-dimensional   theories  by 
considering   a simple  5-dimensional  Abelian model.    After briefly  
discussing how these concepts can be  applied  to the SM in Section 3, 
we turn our attention to the phenomenological aspects of the models of 
our  interest  in  Section 4.  Because of the limited space, technical 
details  are omitted in this note.   A complete discussion, along with 
detailed  analytic  results  and references,  is   given  in our paper 
in~\cite{MPR}. Section 5 summarizes our numerical results and presents 
our conclusions.

\setcounter{equation}{0}
\section{5-Dimensional Abelian Models}\label{5DQED}

As a  starting point, let  us consider the Lagrangian of 5-dimensional
Quantum Electrodynamics (5D-QED) given by
\begin{equation}
\label{freelagrangian}
\la (x, y) \, = \, - \frac{1}{4} F_{M N} (x, y) F^{M N} (x, y) \:
+\: \la_{\mr{GF}}(x,y)\:,
\end{equation}
where $F_{M N}$ denotes the  5-dimensional field strength tensor,  and
$\la_{\mr{GF}}(x,y)$ is the gauge-fixing term. Our notation is: $M,N =
0,1,2,3,5$; $\mu,\nu = 0,1,2,3$; $x = (x^0,\vec{x})$; and $y = x^5$.

In  the absence  of  the  gauge-fixing  and  ghost  terms, the  5D-QED
Lagrangian  is  invariant  under   U(1)  gauge  transformations.    To
compactify the theory on an $S^1/Z_2$ orbifold,   we  demand  for  the 
fields to satisfy equalities like
\begin{equation}
\label{fieldconstraints}
\begin{split}
A_{\mu}(x,y) \, & = \, A_{\mu}(x,y + 2 \pi R)\,, \\ 
A_{\mu}(x,y) \, & = \, A_{\mu}(x, - y)\,.
\end{split}
\end{equation}
The  field $A_\mu (x,y)$ is   taken to be even   under $Z_2$, so as to
embed conventional QED with a massless photon into  our 5D-QED.  Then,
the reflection properties of the $A_5 (x,y)$ field with respect to $y$
are dictated by gauge invariance, i.e.\ $A_5 (x,y) = - A_5 (x,-y)$.

Given (\ref{fieldconstraints}), we  can expand  the fields in  Fourier
series, where  the Fourier  coefficients, denoted $A_{(n)}^{\mu} (x)$,
are  the so-called KK modes.     Integrating out the $y$~dimension  we
obtain the  effective 4-dimensional Lagrangian including massless QED.
The  other  terms  describe   two  infinite  towers of  massive vector
excitations   $A^{\mu}_{(n)}$  and (pseudo)-scalar modes $A^{5}_{(n)}$
that mix   with  each  other,   for  $n  \ge   1$.  The  scalar  modes
$A^{5}_{(n)}$  play  the r\^ole of the  would-be  Goldstone modes in a
non-linear  realization  of  an Abelian   Higgs model,   in which  the
corresponding Higgs fields are taken to be infinitely massive.

The above  observation motivates us to  seek for  a higher-dimensional
generalization of  't-Hooft's  gauge-fixing condition. We   choose the
following generalized $R_\xi$ gauge \cite{MPR,GNN}:
\begin{equation}
\label{gengaugefixterm}
\la_{\mr{GF}}(x, y)\ =\ -\, \frac{1}{2 \xi} (\partial^{\mu} A_{\mu}
\: - \: \xi \, \partial_5 A_5)^2 \, .
\end{equation}
Upon integration over the  extra dimension, all mixing terms disappear
and the Lagrangian describes  QED accompanied  by  a tower  of massive
gauge   bosons  $A_{(n)}^{\mu}$ and  the  respective  Goldstone  modes
$A_{(n)  5}$.  The limit $\xi  \to  \infty$  corresponds to  the usual
unitary  gauge~\cite{PS,DMN}.  Thus, for a simple  model, we have seen
how starting  from   a non-covariant higher-dimensional   gauge-fixing
condition, we can arrive at  the known covariant 4-dimensional $R_\xi$
gauge after compactification.

This quantization procedure can  be  successfully applied to  theories
that include Higgs  and gauge bosons living  in the bulk and/or on the
brane~\cite{GGH}.  A brane    Higgs induces mixing  terms between  the
Fourier modes. The  KK mass eigenstates,   found by diagonalizing  the
mass   matrix,  have slightly shifted  masses   and couplings to brane
fermions compared to the Fourier modes.

\setcounter{equation}{0}
\section{5-Dimensional Extensions of the Standard Model}
\label{MinimalExtensions}

The   ideas   introduced   in   \sect{5DQED}   can be generalized  for 
non-Abelian  theories.    As a   new feature,  the self-interaction of 
gauge-bosons   in non-Abelian   theories leads to self-interactions of 
the KK modes which   are restricted by selection rules reflecting  the 
$S^1/Z_2$ structure of the extra dimension.

For spontaneous symmetry-breaking theories, such as the Standard Model
(SM), the  existence  of   new  compact dimensions opens  up   several
possibilities in connection with  the SU(2)$_L\otimes $U(1)$_Y$ gauge
structure.  For example, the   SU(2)$_L$ and U(1)$_Y$ gauge  fields do
not necessarily need to propagate both in the extra dimension.  Such a
realization  may  be encountered  within specific  stringy frameworks,
where  one  of  the  gauge  groups   is  effectively confined  on  the
boundaries of the $S^1/Z_2$ orbifold~\cite{AB}.

However, in  the most frequently  investigated scenario, SU(2)$_L$ and
U(1)$_Y$ gauge   fields  live in   the  bulk of  the   extra dimension
(bulk-bulk model).  In this  case, for generality,  we will consider a
2-doublet  Higgs model, where one  Higgs field propagates in the fifth
dimension, while the  other  one is  localized.  The phenomenology  of
this model is  influenced by the  vacuum expectation values  $v_1$ and
$v_2$, or equivalently by $\tan \beta = v_2 / v_1$ and  $v^2 = v^2_1 +
v^2_2$.

An even  more minimal  5-dimensional extension of  electroweak physics
constitutes a model in which only the SU(2)$_L$-sector feels the extra
dimension  while  the  U(1)$_Y$  gauge  field is  localized  at  $y=0$
(bulk-brane model). In  this case, the Higgs field  being charged with
respect to both gauge groups has  to be localized at $y=0$ in order to
preserve gauge invariance of  the (classical) Lagrangian. For the same
reason, a bulk Higgs is forbidden in the third possible model in which
SU(2)$_L$  is  localized  while   U(1)$_Y$  propagates  in  the  fifth
dimension (brane-bulk model).

In all these minimal 5-dimensional extensions of the SM we assume that
the  SM  fermions are  localized  at  the $y=0$   fixed point  of  the
$S^1/Z_2$ orbifold. All the KK modes of a bulk field couple to a brane
fermion. Because  the KK  mass  eigenmodes generally differ   from the
Fourier modes, their  couplings to fermions  have to be calculated for
each model individually.

\setcounter{equation}{0}
\section{Effects on Electroweak Observables}
\label{Phenomenology}

In this section, we will concentrate on  the phenomenology and present
bounds   on the  compactification scale      $M =  1/R$ of     minimal
5-dimensional  extensions of the SM   calculated by analyzing a  large
number  of high precision electroweak   observables.  We relate the SM
prediction  $\mathcal{O}^{\mr{SM}}$~\cite{PDG,EWWG} for  an observable  
to the prediction  $\mathcal{O}^{\mr{HDSM}}$  for the same  observable 
obtained in  the higher-dimensional SM under  investigation through
\begin{equation}
\label{generalformofpredictions}
{\cal O}^{\rm HDSM} \ =\ {\cal O}^{\rm SM} \, \big( 1\: +\: 
\Delta^{\rm HDSM}_{\cal O} \big)\, .
\end{equation}
Here, $\Delta^{\rm HDSM}_{\cal O}$ is the tree-level modification of a
given observable ${\cal O}$  from its SM value  due to the presence of
one  extra  dimension.    In  order  to  enable  a  direct  comparison
of  our  predictions  with  the  precision  data~\cite{PDG,EWWG},   we
include SM radiative corrections to~$\mathcal{O}^{\mr{SM}}$.  However,
we neglect SM-  as  well  as  KK-loop contributions  to   $\Delta^{\rm
HDSM}_{\cal O}$ as higher order effects.

As input SM parameters for our  theoretical predictions, we choose the
most accurately measured  ones,  namely the $Z$-boson mass  $M_Z$, the
electromagnetic   fine   structure constant~$\alpha$    and the  Fermi
constant $G_F$.  While $\alpha$ is  not  affected in the models  under
study,  $M_{Z}$  and $G_F$  generally deviate  from their SM form when 
expressed  in  terms  of  couplings  and  VEV's.  To  first  order  in 
$X=\frac{1}{3}\pi^2       m^2_Z R^2$,    $M_{Z}$  and  $G_F$   may  be 
parameterized as
\begin{equation}
\label{deltazdef}
M_{Z} \  =\ M^{\mr{SM}}_{Z} \, \lr{1 \: + \: \Delta_{Z}\,X}\,, \qquad
G_F \ = \ G^{\mr{SM}}_F \, \lr{ 1\: +\: \Delta_G\, X}\,,
\end{equation}
where $\Delta_{Z}$  and $\Delta_G$ are model-dependent parameters. For
example, one finds
\begin{equation}
\Delta_{Z} = \big\{ \ - \, \frac{1}{2} \,  \sin^4\beta\,,\ - 
\, \frac{1}{2} \, \sin^2 \hat{\theta}_W\,,\ - \, \frac{1}{2} \, 
\cos^2\hat{\theta}_W\, \big\}\,.
\end{equation} 
for  the bulk-bulk, brane-bulk  and bulk-brane models, with respect to
the SU(2)$_L$ and U(1)$_Y$ gauge groups.

The  relation between the weak  mixing  angle $\theta_W$ and the input
variables is also affected by the fifth dimension. Hence, it is useful
to  define an   effective  mixing angle  $\hat{\theta}_W$, which still
fulfills the tree-level relation
\begin{equation}
\label{definitionequationfonormalizedthetaw}
G_F \ = \ \frac{\pi \alpha}{\sqrt{2} \sin^2 \hat{\theta}_W \, \cos^2
\hat{\theta}_W \, M_{Z}^2}
\end{equation}
of the Standard Model,  and relate it to $\theta_W$ by 
$\sin^2 \hat{\theta}_W  =  \sin^2 \theta_W  \, \lr{  1\: +\:
\Delta_{\theta}\,X }$.

For the tree-level calculation of $\Delta^{\rm  HDSM}_{\cal O}$, it is
necessary to consider the mixing  effect of the  Fourier modes on  the
masses of   the Standard-Model  gauge  bosons   as well as    on their
couplings to  fermions. All the  encountered shifts can be expanded in
powers of $X$  and are calculated  to  first order. For the  precision
measurement at the  $Z$ pole  or at lower  energies these  effects are
dominant.  For cross  sections at  LEP2  energies, the dominant higher
dimensional contributions stem  from the interference  of the Standard
Model with virtual KK modes which roughly scales like $s/M^2$.

Within   the   framework outlined   above,  we   compute  $\Delta^{\rm
HDSM}_{\cal     O}$    for    an    extensive    list    of  precision 
observables~\cite{MPR}.   In   addition,   we  consider   fermion-pair 
production   at   LEP2   \cite{MPR2}.   Employing   the   results   of  
$\Delta^{\mr{HDSM}}_{\mathcal{O}}$ and calculating all the observables  
considered        in        our       analysis         by       virtue 
of~\equ{generalformofpredictions}, we  confront these predictions with
the  respective experimental values  and  calculate the  corresponding
$\chi^2 (X)$ where it  is  important to include correlations   between
some of the observables. The bounds on $X$ can be derived by requiring
$\chi^2 (X) - \chi^2_{\mr{min}} < n^2 $ for $X$  being not excluded at
the $n \,  \sigma$ confidence level.  Here, $\chi^2_{\mr{min}}$ is the
minimal $\chi^2$ in   the physical region  $X  \ge 0$.  Using slightly
different definitions for the   bounds does not lead to  significantly
different results.

Table~\ref{bounds} summarizes the lower bounds on the compactification
scale $M  = 1/R$ coming from  different observables. For the bulk-bulk
model we consider the two extreme cases, a pure  bulk Higgs and a pure
brane Higgs.

\begin{table}
\begin{center}
\begin{tabular}{c|c|c|c|c} 
\hline\hline 
& \pb{2.8}{-0.}{ SU(2)$_L$-brane, \\ U(1)$_Y$-bulk }
& \pb{2.8}{-0.}{ SU(2)$_L$-bulk, \\ U(1)$_Y$-brane }
& \pb{2.8}{-0.}{SU(2)$_L$-bulk, \\ U(1)$_Y$-bulk \\ (brane Higgs) }
& \pb{2.8}{-0.}{ SU(2)$_L$-bulk, \\ U(1)$_Y$-bulk \\ (bulk Higgs) } \\ \hline \hline  
prec. obs. 
& \pb{2}{-0.}{4.2} & \pb{2}{0.}{2.9}
& \pb{2}{-0.}{4.6} & \pb{2}{0.}{4.6} \\ \hline
$\mu^+ \mu^-$ & \pb{2}{0.}{ 2.0 } & \pb{2}{0.}{ 1.5 } 
& \pb{2}{0.}{ 2.5 } & \pb{2}{0.}{ 2.5 } \\ \hline
$\tau^+ \tau^-$ & \pb{2}{0.}{ 2.0 } & \pb{2}{0.}{ 1.5 } 
& \pb{2}{0.}{ 2.5 } & \pb{2}{0.}{ 2.5 } \\ \hline 
hadrons & \pb{2}{0.}{ 2.6 } & \pb{2}{0.}{ 4.7 } 
& \pb{2}{0.}{ 5.4 } & \pb{2}{0.}{ 5.8 } \\ \hline 
$e^+ e^-$ & \pb{2}{0.}{ 3.0 } & \pb{2}{0.}{ 2.0 } 
& \pb{2}{0.}{ 3.6 } & \pb{2}{0.}{ 3.5 } \\ \hline 
combined  & \pb{2}{0.}{ 4.7 } & \pb{2}{0.}{ 4.3 } 
                 & \pb{2}{0.}{ 6.1 } & \pb{2}{0.}{ 6.4 } \\ \hline
\end{tabular}
\end{center}
\mycaption{\label{bounds} Bounds on the compactification scale at the 
$2   \sigma$ confidence   level  from  precision observables   and the
different fermion-pair production channels at LEP2. }
\end{table}

\setcounter{equation}{0}
\section{Discussion and Conclusions}

\indent

By performing $\chi^2$-tests, we obtain different sensitivities to the
compactification radius $R$ for  the three models under consideration:
(i)~the SU(2)$_L\otimes$U(1)$_Y$-bulk model, where all SM gauge bosons
are bulk fields; (ii)  the SU(2)$_L$-brane, U(1)$_Y$-bulk model, where
only  the SU(2)$_L$ fields are  restricted to the brane, and (iii)~the
SU(2)$_L$-bulk, U(1)$_Y$-brane  model, where only  the U(1)$_Y$  gauge
field is  confined to the  brane.  The strongest  bounds hold  for the
often-discussed bulk-bulk model no matter if the Higgs boson is living
in the  bulk or on the brane.   For the bulk-brane  models, we observe
that the combined bounds on $1/R$ are reduced by roughly 20 to 30\%.

The lower limits on the compactification  scale derived by the present
global analysis indicate that   resonant  production of the  first  KK
state may be at the edge of the LHC reach, at which heavy KK masses up
to  6--7~TeV~\cite{AB,RW} might be explored. One  probably will not be
able to probe resonant effects  originating from the second KK  state,
and so more phenomenological work has  to be done to differentiate the
model from other 4-dimensional new-physics scenaria.

In  addition, we have paid  special attention to consistently quantize
the  higher-dimen\-sional models in   the generalized $R_\xi$  gauges.
Specifically,   we  have  been   able    to identify the   appropriate
higher-dimensional gauge-fixing  conditions which should be imposed on
the theories  so as to  yield the known  $R_\xi$ gauge after the fifth
dimension has been integrated out \cite{MPR,DCH}.

\subsection*{Acknowledgements}
This  work was supported  by the  Bundesministerium f\"ur  Bildung and
Forschung (BMBF,  Bonn, Germany) under the  contract number
05HT1WWA2.


\begin{thebibliography}{99}

\bibitem{review} For a review, see e.g., M.B. Green, J.H. Schwarz and
  E. Witten, ``Superstring Theory,'' (Cambridge, Cambridge University
  Press, 1987).

\bibitem{IA} I. Antoniadis, Phys.\ Lett.\ {\bf B246} (1990) 377.

\bibitem{JL} J.D. Lykken, Phys.\ Rev.\ {\bf D54} (1996) 3693.
  
\bibitem{EW} E. Witten, Nucl.\ Phys.\ {\bf B471} (1996) 135; P.
  Ho$\check{{\rm r}}$ava and E. Witten, Nucl.\ Phys.\ {\bf B460}
  (1996) 506; Nucl.\ Phys.\ {\bf B475} (1996) 94.

\bibitem{ADS} N. Arkani-Hamed, S. Dimopoulos, G. Dvali, Phys.\ Lett.\ 
  {\bf B429} (1998) 263; I. Antoniadis, N. Arkani-Hamed, S. Dimopoulos
  and G. Dvali, Phys.\ Lett.\ {\bf B436} (1998) 257; N. Arkani-Hamed,
  S. Dimopoulos and G. Dvali, Phys.\ Rev.\ {\bf D59} (1999) 086004.

\bibitem{DDG} K.R. Dienes, E.  Dudas and T. Gherghetta, Phys.\ Lett.\
  {\bf B436} (1998) 55; Nucl.\ Phys.\ {\bf B537} (1999) 47.
    
\bibitem{AB} I. Antoniadis and K. Benakli, Int.\ J. Mod.\ Phys.\ {\bf
    A15} (2000) 4237.
  
\bibitem{NY} P.  Nath and M.  Yamaguchi, Phys.\ Rev.\ {\bf D60} (1999)
  116006; Phys.\ Lett.\ {\bf B466} (1999) 100.
  
\bibitem{WJM} W.J. Marciano, Phys.\ Rev.\ {\bf D60} (1999) 093006; M.
  Masip and A. Pomarol, Phys.\ Rev.\ {\bf D60} (1999) 096005.
  
\bibitem{CCDG} R. Casalbuoni, S. De Curtis,  D. Dominici and R. Gatto,
Phys.\ Lett.\ {\bf B462} (1999)  48; C.~Carone, Phys.\ Rev.\ {\bf D61}
(2000) 015008.

\bibitem{DPQ2} A. Delgado, A. Pomarol and M. Quiros, JHEP {\bf 0001}
  (2000) 030.
    
\bibitem{RW} T. Rizzo and J. Wells, Phys.\ Rev.\ {\bf D61} (2000)
  016007; A. Strumia, Phys.\ Lett.\ {\bf B466} (1999) 107.
  
\bibitem{DPQ1} A. Delgado, A. Pomarol and M. Quiros, Phys.\ Rev.\
  {\bf D60} (1999) 095008.
  
\bibitem{CL} K. Cheung and G. Landsberg, 
Phys. Rev.~{\bf D65} (2002) 076003.

\bibitem{MPR} A.  M\"uck, A.  Pilaftsis and  R.  R\"uckl, Phys.\ Rev.\
{\bf D65} (2002) 085037.

\bibitem{GNN} D.M. Ghilencea, S. Groot Nibbelink and H.P. Nilles, 
Nucl.\ Phys.\ {\bf B619} (2001) 385.

\bibitem{PS} J. Papavassiliou and A. Santamaria,
Phys.\ Rev.\ {\bf D63} (2001) 125014.

\bibitem{DMN}  D. Dicus, C. McMullen  and S.  Nandi, Phys.\ Rev.\ {\bf
D65} (2002) 076007.
  
\bibitem{GGH} For example, see H. Georgi, A.K. Grant and G. Hailu,
Phys.\ Lett.\ {\bf B506} (2001) 207; M.~Carena, T. Tait and
C.E.M. Wagner, hep-ph/0207056.

\bibitem{PDG} K.~Hagiwara {\it et al.}  [Particle Data Group
  Collaboration], Phys.\ Rev.\ {\bf D66} (2002) 010001.

\bibitem{EWWG}  The LEP Collaborations ALEPH, DELPHI, L3, OPAL, the LEP Electroweak 
  Working Group and the SLD Heavy Flavor and Electroweak Groups, hep-ex/0112021.

\bibitem{MPR2}  A. M\"uck,  A.  Pilaftsis  and  R.   R\"uckl: work  in
preparation.

\bibitem{DCH} R.S. Chivukula, D.A. Dicus and H.-J. He, 
Phys. \ Lett. \ {\bf B525} (2002) 175.

\end{thebibliography}
\end{document}